\documentclass[twocolumn,showpacs,preprintnumbers,amsmath,amssymb,prl]{revtex4}

\usepackage{graphicx}
\usepackage{dcolumn}
\usepackage{epsf}

\begin{document}

\title{Can antiferromagnetism and superconductivity coexist in the high-field paramagnetic superconductor Nd(O,F)FeAs?}

\author{C. Tarantini$^{1}$, A. Gurevich$^{1}$, D.C. Larbalestier$^{1}$, Z.A. Ren$^{2}$, X.L. Dong$^{2}$, W. Lu$^{2}$, Z.X. Zhao$^{2}$.}
\affiliation{$^{1}$National High Magnetic Field Laboratory, Florida State University, Tallahassee, Florida, 32310 USA, \\
$^{2}$National Laboratory for Superconductivity, Institute of Physics and Beijing National Laboratory
for Condensed Matter Physics, Chinese Academy of Sciences, P.O. Box 603, Beijing 100190, P.R. China.}

\date{\today}
\begin{abstract}

We present measurements of the temperature and field dependencies of
the magnetization $M(T,H)$ of Nd(O$_{0.89}$F$_{0.11}$)FeAs at fields
up to 33T, which show that superconductivity with the critical
temperature $T_c\approx 51$K cannot coexist with antiferromagnetic
ordering. Although $M(T,H)$ at $55<T<140$K exhibits a clear
Curie-Weiss temperature dependence corresponding to the Neel
temperature $T_N\approx 11-12$K, the behavior of $M(T,H)$ below
$T_c$ is only consistent with either paramagnetism of weakly
interacting magnetic moments or a spin glass state. We suggest that
the anomalous magnetic behavior of an unusual high-field
paramagnetic superconductor Nd(O$_{1-x}$F$_x$)FeAs is mostly
determined by the magnetic Nd ions.

\end{abstract}
\pacs{\bf 74.20.De, 74.20.Hi, 74.60.-w}

\maketitle

The recently discovered superconductivity in the family of doped
oxypnictides \cite{as} exhibits an unusual coexistence of the rather
high critical temperature $T_c\simeq 26-57$ K with strong
antiferromagnetic (AF) correlations and apparent magnetic ordering.
Indeed, neutron \cite{neutr1,neutr2}, $\mu SR$ \cite{musr1,musr2},
NMR \cite{nmr1,nmr2} and Mossbauer \cite{mosb1,mosb2} measurements
have revealed complex magnetic structures including a spin density
wave AF instability at $T_N\simeq 140-150$ K accompanied by the
resistivity drop and monoclinic distortions of the tetragonal
lattice of the parent semimetal compound \cite{sdw}. Doping with F
or O reduces $T_N$ and seemingly eliminates manifestations of the AF
ordering in the temperature dependence of the resistivity $\rho(T)$
in oxypnictides with the optimum $T_c$ \cite{wen,ornl}. This
behavior poses a fundamental question if the AF ordering is merely a
consequence of the nesting of the Fermi surface in the parent
semimetal or it does play a major role in superconductivity. Recent
theoretical models have proposed that magnetic correlations are
instrumental for the unconventional superconductivity in
oxypnictides \cite{sdw1,sdw2,sdw3,sdw4,sdw5}, which may also be
important for understanding their extremely high upper critical
fields $H_{c2}$ \cite{hc2a,hc2b}.

In this Letter we address the specific form in which the magnetic
correlations manifest themselves both in the normal and
superconducting state of the Nd(O$_{1-x}$F$_x$)FeAs using high
magnetic fields $H$ up to 33T capable of destroying any global AF
state with $T_N<\mu_BH/k_B \simeq 22$ K, where $\mu_B$ is the Bohr
magneton. Our measurements of the magnetization $M(T,H)$ in a wide
range of fields $0<H<33$ T and temperatures $5<T<140$ K have shown a
strong paramagnetic contribution in $M(T,H)$ both at $T>T_c$ and
$T<T_c$. In the normal state $M(T)$ exhibits a clear Curie-Weiss
temperature dependence with an apparent Neel temperature $T_N\simeq
11-12$ K, which might suggest a coexistence of superconductivity and
antiferromagnetism. However, our data at $T<T_c$ show instead that
$M(T,H)$ in the superconducting state is inconsistent with a
long-range AF ordering, indicating that superconductivity
effectively cuts off the AF interaction between magnetic moments
evident at $T>T_c$. We suggest that Nd(O$_{1-x}$F$_x$)FeAs may be
regarded as a {\it paramagnetic} superconductor with unusual
magnetic behavior different not only from conventional
superconductors but also from La(O$_{1-x}$F$_x$)FeAs oxypnictides.
This difference may result from a large magnetic moment
$\mu=3.62\mu_B$ of a free Nd$^{3+}$ ion ($\mu= 3.58\mu_B$ for
Pr$^{3+}$), as compared to $\mu=0$ for La$^{3+}$ and $\mu=0.8\mu_B$
for Sm$^{3+}$ \cite{mag}. Thus, $m(T,H)$ of Nd(O,F)FeAs is mostly
dominated by the Curie-Weiss behavior of Nd$^{3+}$, while
contributions from Fe$^{3+}$ with $\mu=0.25-0.35 \mu_B$
\cite{neutr1,neutr2} and temperature-independent van-Vleck and Pauli
paramagnetism become negligible.

The Nd(O$_{0.89}$F$_{0.11}$)FeAs polycrystalline samples were made
as described in Ref. \cite{sample}. The sample with approximate
dimensions of $1.4\times 1\times 0.5$ mm$^3$, volume $V\approx 0.7$
mm$^{-3}$ and weight $\approx 4.5$ mg had $\rho(T_c)\approx  0.53$
m$\Omega$cm, $T_c \approx 51$ K, and about $90\%$ of the theoretical
density, $7.213$ g/cm$^3$. The magnetic susceptibility $\chi(T,H)$
and magnetic moment $M(T,H)V$ were measured by SQUID magnetometer,
Vibrating Sample Magnetometer (VSM) in a 14T superconducting magnet
and a resistive 33T magnet at NHMFL. Shown in Figs. 1 and 2 are the
zero field cooled (ZFC) and field cooled (FC) $M(T)$ curves, which
change radically upon increasing field. For $H=10$mT, $M(T)$
exhibits a standard superconducting behavior with the diamagnetic
onset close to $T_c$. However, increasing field to a rather modest
value of 0.5T (given $H_{c2}(0)\sim 50-100 T$ for oxypnictides
\cite{hc2a})), shrinks the irreversible diamagnetic loop, while
significantly increasing the paramagnetic component of $M(H)$.
Further increase of $H$ to 1.5T, practically eliminates the
difference between the ZFC and FC curves, both exhibiting a strong
paramagnetic behavior. Fig. 2 shows in more detail how this
field-induced superconducting to paramagnetic transition develops.

    \begin{figure}                  
    \epsfxsize= 0.7\hsize
    \centerline{
    \vbox{
    \epsffile{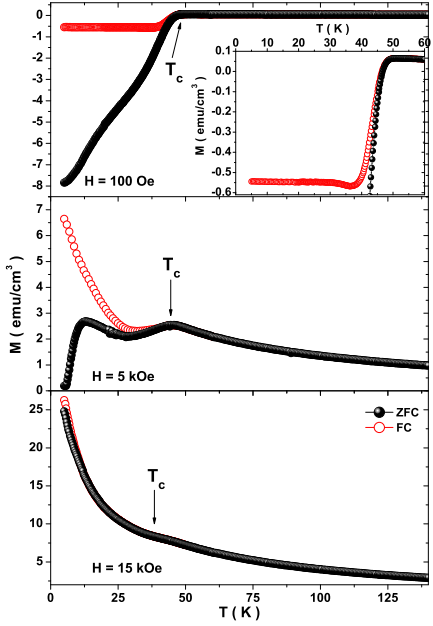}
    }}
    \caption{ZFC (black full dots) and FC (red open dots) $M(T)$
    at: 0.1, 5 and 15 kOe. Inset shows FC $M(T)$ at 0.1 kOe.}
    \label{Fig.1}
    \end{figure}

    \begin{figure}                  
    \epsfxsize= 0.75\hsize
    \centerline{
    \vbox{
    \epsffile{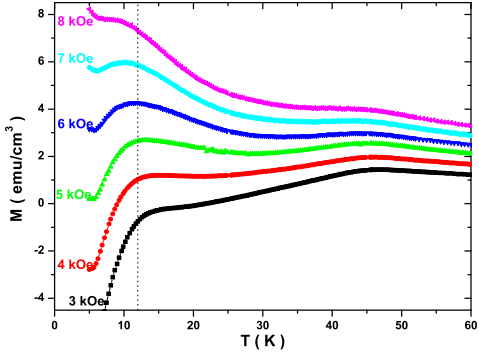}
    }}
    \caption{ZFC magnetization at different fields.
    The dash line shows $T_m$ extracted from the data in Fig.3. }
    \label{Fig.1}
    \end{figure}

To gain a further insight into the nature of the paramagnetic signal
not masked by superconductivity, $M(T,H)$ between 55 and 140 K was
measured by a SQUID magnetometer. The results shown in Fig. 3
unambiguously indicate the AF Curie-Weiss behavior, $M^{-1}\propto
T+T_m$ with $T_m\approx 11-12$K being practically independent of $H$
at $0<B<5$ T. This may indicate AF ordering at the Neel temperature
$T_N\simeq T_m$, {\it below} the superconducting transition and a
coexistence of superconductivity and antiferromagnetism in a variety
of forms, for example, a quantum critical point as a function of
doping. To see if this is indeed the case, we measured $M(T,H)$
below $T_c$ at very high magnetic fields, which enabled us to
suppress the irreversible magnetization due to pinned vortices and
to probe $M(T,H)$ in the superconducting state at fields $H>
T_N/\mu$ sufficient to destroy the global AF state.

    \begin{figure}                  
    \epsfxsize= 0.75\hsize
    \centerline{
    \vbox{
    \epsffile{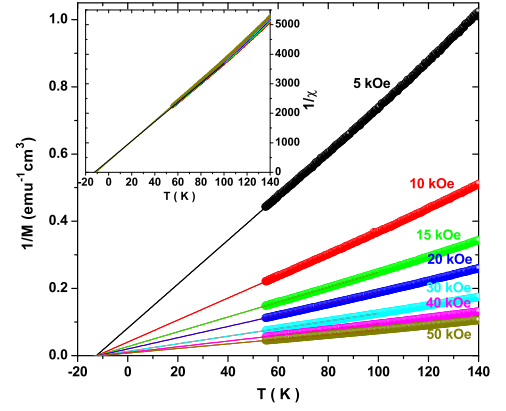}
    }}
    \caption{The AF Curie-Weiss linear temperature dependence of $M^{-1}(T,H)\propto T+T_m$ in the
    normal state. The linear extrapolations defines $T_m \approx 11-12$K.
    The inset shows the collapse of the inverse susceptibilities $\chi^{-1}(T,H)=H/M$ for different fields
    $0<H<5$ T onto a single straight line.}
    \label{Fig.3}
    \end{figure}

Shown in Fig. 4a is the $M(H)$ loop at 4.2 K for the field
perpendicular to the broader sample face. One can clearly see a
superconducting hysteresis loop on top of a strongly paramagnetic,
field-dependent background. Taking the mean value of the two
branches of $m(B)$ as the paramagnetic magnetization, $M_p
=(M_\uparrow + M_\downarrow)/2$ plus the small reversible
magnetization of the vortex lattice, we define the magnetization
$\Delta M = M_\uparrow - M_\downarrow$ due to the critical state of
pinned vortices. Using the Bean model, we estimate the critical
current density $J_c(4.2K) \sim 10$ kA/cm$^2$ at self field. $M(H)$
in Fig. 4a increases with field and tends to level off at higher H.
To probe the region of higher fields, $M(H,T)$ was measured in a 33T
resistive magnet. The results presented in Fig. 4b, show that
$M(4.2K,H)$ does saturate, while $M(T,H)$ at higher temperatures
increases much slower as H increases. The width of superconducting
loop decreases as T and H increase, consistent with the usual
behavior of $J_c(H,T)$.

    \begin{figure}                  
    \epsfxsize= 0.8\hsize
    \centerline{
    \vbox{
    \epsffile{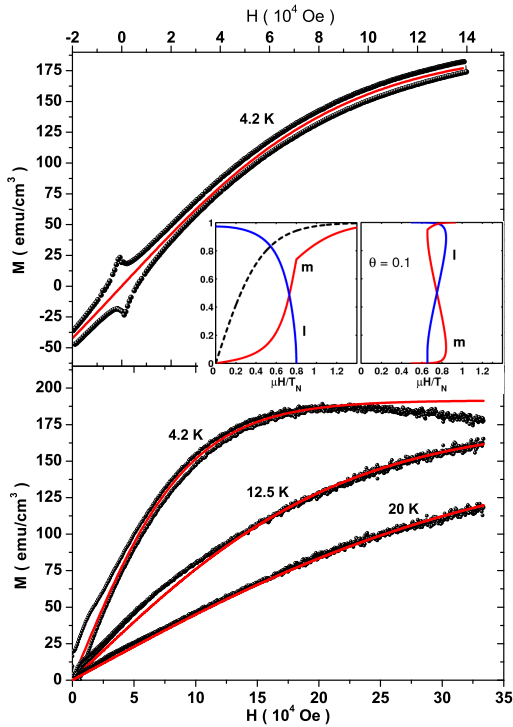}
    }}
    \caption{(a) Low-field $M(H)$ loop measured by VSM at 4.2 K.
    The red line shows $M=M_0\tanh(\mu B/T)$ with $B=H+4\pi M$, $M_0=\mu n$
    for $\mu=0.69\mu_B$, $M_0V= 0.135$ emu and $n=3\times 10^{22}$ cm$^{-3}$. Inset shows
    $m(H)$ and $l(H)$ calculated from Eqs. (\ref{z})-(\ref{par}) for $\gamma=0.5$, and $m=\tanh(h)$ (dashed line).
    Here the left panel shows the second order transition for
    $\theta = 5/11$, and the right panel shows the first order metamagnetic transition for $\theta = 0.1$.
    (b) High-field $M(H)$ loops at different T. The
    red lines show $M=M_0\tanh(\mu B/T)$ with $\mu=0.67\mu_B$, $M_0V= 0.133$ emu, $n=3.1\times 10^{22}$ cm$^{-3}$ (4.2K);
    $\mu=0.85\mu_B$, $M_0V= 0.123$ emu, $n=2.25\times 10^{22}$ cm$^{-3}$ (12.5 K); $\mu=0.85\mu_B$,
    $M_0V= 0.112$ emu, $n=2.05\times 10^{22}$ cm$^{-3}$ (20 K). }
    \label{Fig.4}
    \end{figure}

To interpret our data we used the mean-field
AF theory in which two magnetic sublattices with the magnetizations
$M_1$ and $M_2$ are controlled by the molecular fields
$H_1=H+aM_1-bM_2$ and $H_2=H+aM_2-bM_1$, where $a$ and $b$ quantify
the exchange interaction with nearest neighbors of the same spin
sublattice (a) and the sublattice with the opposite spins (b)
\cite{mag}. The self-consistency equations, $2M_i=M_0B_J(\mu
H_i/T)$, $i=1,2$ determine the total magnetization,
$M(T,H)=M_1+M_2$, and the AF order parameter $L(T,H)=M_1-M_2$ where
$M_0=n\mu$, $\mu$ is an elementary magnetic moment, $n$ is their
total density,
$B_J(z)=[(2J+1)/2J]\coth[(2J+1)z/2J]-(1/2J)\coth(z/2J)$ is the
Brillouin function, and $J$ is the maximum magnetic quantum number.
Expansion of $B_J(\mu B_i/T)$ in $\mu H_i/T$ at $T\gg T_N$ gives the
linear temperature dependence for 
$1/\chi(T,H)=H/M$:
    \begin{equation}
    \frac{\chi(T_*)}{\chi(T,H)}=\frac{T+T_m(H)}{T_*+T_m(H)},
    \label{chi}
    \end{equation}
Here $T_m=(b-a)M_0(J+1)/6J$, $T_*=2T_m$, $T_m(B)=[1-c_J(\mu
H/T)^2]T_m$, $c_J=(2J^2+2J+1)/10J^2$, and $\chi(T_*)=9JT_m/(J+1)\mu
M_0$, $T_m = \gamma T_N$ where $\gamma=(b-a)/(b+a)<1$. Eq.
(\ref{chi}) describes the data in Fig. 3 well, allowing us to
estimate the density n of magnetic moments $\mu$, which can provide
the observed $\chi$. Taking $\chi_0=n\mu^2/3T_m=1/395$ from the
inset of Fig. 3, $T_m=12$ K, and $\mu=3.62\mu_B$ for Nd$^{3+}$
\cite{mag}, we obtain $n=3T_m\chi_0/\mu^2\simeq 1.5\times10^{22}$ cm$^{-3}$,
close to the density of Nd atoms. Since the strong paramagnetism
above $T_c$ is only consistent with an atomic density of magnetic
moments $\mu\simeq 3-4\mu_B$, Fe$^{3+}$ with
much smaller $\mu\simeq 0.25-0.35\mu_B$ \cite{neutr1,neutr2} can
only add a few percent to $\chi\propto \mu^2$. The crystalline field
splits the degenerate orbital term of Nd$^{3+}$ \cite{mag,split}
into terms with spin moments $\mu$ smaller than $\mu$ of the free 
Nd$^{3+}$ but with $n$ greater than the atomic density.

We consider breakdown of AF by strong fields using a simple model,
in which only spin degrees of freedom
contribute to $M$, so the mean-field equations, $2M_1=M_0\tanh(\mu H_1/T)$ and
$2M_2=M_0\tanh(\mu H_2/T)$ can be reduced to the following
parametric equations
    \begin{eqnarray}
    z=\cosh^{-1}[(2/\theta s)\sinh s - \cosh s],
    \label{z} \\
    m=\sinh z/(\cosh z + \cosh s), \\
    \label{x}
    l=\theta s/2,\qquad h = z\theta/2+\gamma x,
    \label{yh}
    \end{eqnarray}
which define the dimensionless magnetization $m=M/M_0$ and the order
parameter $l=L/M_0$ as functions of the dimensionless field $h=\mu
H/T_N$ and temperature $\theta = T/T_N$. Here the parameter $s$ runs
from $0$ to $2/\theta$, and $T_N=M_0(b+a)/2$. For $T<T_N$, the AF order
parameter $L(H,T)$ decreases as $H$ increases, vanishing at the
second order phase transition field $H=H_p$:
    \begin{equation}
    H_p=[\gamma\sqrt{1-\theta}+\theta\cosh^{-1}(1/\sqrt{\theta})]T_N/\mu,
    \label{hp}
    \end{equation}
at which $M(H_p)=M_0\sqrt{1-\theta}$. For $H>H_p$, the magnetization in the paramagnetic phase is
described by
    \begin{equation}
    m=\tanh[(h-\gamma x)/\theta]
    \label{par}
    \end{equation}
Shown in the inset of Fig. 4 are $m(h)$ and $l(h)$ curves calculated
from Eqs. (\ref{z})-(\ref{par}) for high $(\theta = 5/11)$ and low
$(\theta = 0.1)$ temperatures. At high temperatures $m(h)$ has an
upward curvature at $H<H_p$ and the downward curvature above the
second order phase transition field $H_p$ at which $L(H)$ vanishes.
At low temperatures, $m(h)$ and $l(h)$ become multivalued, indicating
hysteretic first order metamagnetic
transitions due to spin flip in the sublattice with $m$ antiparallel to $H$.

One can immediately see a very different behavior of the mean-field
$M(H)$ as compared to the observed $M_p(H)$. Indeed, all $M_p(H)$
curves exhibit downward curvatures and can be described well by the
Brillouin function $M=M_0\tanh(\mu H/T)$ with $M_0=\mu n$,
$\mu=(0.67-0.85)\mu_B$ and $n=(2-3)\times 10^{22} $cm$^{-3}$ (the
full set of fit parameters is given in the caption of Fig. 4). These
values of $n$  are about 2 times higher than the density of Nd$^{3+}$ 
in Nd(O,F)FeAs, which may indicate splitting of the $^4I_{9/2}$
ground term of Nd$^{3+}$ into different spin levels. Yet $M_p(H)$
shows no long-range AF behavior below $T_c$, otherwise $m(H)$ would
be significantly depressed at $H<H_p$, as depicted in the inset of
Fig. 4. For example, if $T_N=11$K, $\theta = 5/11$, $\mu=0.8\mu_B$,
and $\gamma = 0.5$, Eq. (\ref{hp}) yields the field $H_p\simeq 16$T
required to destroy the AF state. However, $M(H)$ in Fig. 4a
exhibits neither the downward curvature nor metamagnetic transitions
characteristic of the mean-field behavior at $H<H_p$, while the data
in Fig. 4b exhibit no sign of this behavior on a greater magnetic
field scale either. Therefore, the long-range AF ordering below
$T_c$ is inconsistent with our data, in agreement with the
conclusion drawn from neutron scattering measurements on La(O,F)AsF
\cite{neutr1}. At the same time, the Brillouin function for
noninteracting spins gives a good description of the observed
behavior of $M(T,H)$ at $T<T_c$, particularly the saturation of
$M(H)$ at higher H and decreasing the low-field slope of $M(T)\sim
\mu^2nH/T$ as T increases.

In polycrystals $M(T,H)$ should be averaged with respect to easy
magnetization directions in randomly oriented crystallites. For a
uniaxial magnetic anisotropy and uniform distribution of easy
magnetization axes inclined by angles $\theta$ with respect to H, we
have, $\bar{M}=\int_0^{\pi/2} M(\mu
H\cos\theta/T)\cos\theta\sin\theta d\theta$ and $\chi = \chi_\|/3 +
2\chi_\perp/3$, where $\chi_\| = 2(1-l^2)T_N/(a+b)[T+(1-l^2)\gamma
T_N]$ and $\chi_\perp = 1/b$. If the in-plane AF structure is
present, one can expect suppression of the AF state and local
metamagnetic transitions at lower T in the grains for which the
in-plane field component $H\cos\theta$ exceeds $H_p$. Also
out-of-plane spin-flip transitions may occur if the perpendicular
field component $H\sin\theta$ exceeds $H_c=[2K/(\chi_\perp
-\chi_\|)]^{1/2}$, where $K$ is the energy of magnetic anisotropy
\cite{mag}.

Now we discuss possible mechanisms behind the observed effects. The
first candidate would be paramagnetism of Nd$^{3+}$ ions whose
exchange interaction is mediated by conducting electrons on the AsFe
planes. As follows from Fig. 3, this interaction could provide the
AF ordering in the normal state below $T_N= T_m/\gamma\sim 10-20$K
with $\gamma<1$. However, the superconducting transition in the AsFe
planes sandwiched between Nd(O,F) planes occurs at temperatures
$T_c>T_N$ effectively cutting off the 3D RKKY interaction \cite{imp}
of Nd ions from different planes and reducing $T_N$ of the 2D XY
moments of single Nd planes to zero. This may explain why the
simplest Brillouin function describes our data so well. Since the
paramagnetic signal is so strong that the density of the magnetic
moments $n\sim (2-3)\times 10^{22}$ cm$^{-3}$ is comparable to the
atomic density of Nd ions, we can rule out that the behavior of
$M(T,H)$ is controlled by a small fraction of partially reacted
magnetic second phases.

Our high field measurements enable us to set a lower limit for $T_N$
of AF ordered Fe$^{3+}$ sublattice. In this case the longitudinal
susceptibility $\chi_\|$ at low $T$ and $H<H_p$ is dominated by
paramagnetism of Nd ions. However, as $H$ becomes of the order of
$T_N/\mu$ the AF magnetization of the Fe sublattice becomes strongly
nonlinear, exhibiting metamagnetic or spin flop transitions, which
would manifest themselves in jumps on $m(H)$ curves. Since no sign
of such behavior has been observed up to $H_{max}=33$ T, the
long-range AF order can only be consistent with our results if $H_p$
given by Eq. (\ref{hp}) is greater than $H_{max}$, thus $T_N\gg
T_p$, where $H_p(T_p)=H_{max}$.  For $H_{max}=33T$, $\mu =
0.35\mu_B$ \cite{nmr1,nmr2}, $T=20$ K and $\gamma=0.5$, Eq.
(\ref{hp}) gives $T_p\approx 21$ K. However, $T_N$ much greater than
21K would be hard to reconcile with the Curie-Weiss behavior in Fig.
3 with $T_m =\gamma T_N\approx 11-12$ K, unless $\gamma\ll 1$.

Another possibility is a spin-glass state for which $m(H)$ generally
exhibits the behavior similar to that in the left inset in Fig. 4
with a broadened cusp around $H_p$ and possible kinks and hysteretic
jumps \cite{sg}.  Yet the behavior resembling that of
$M(H,4.2K)$ has also been observed on some metallic spin glasses,
although on a much smaller field scale \cite{sg}. Such a spin glass
state could result from Nd ions and weakly coupled AF clusters of
Fe$^{3+}$, so that $\bar{M}=M_0\int_0^\infty F(\mu_i)\tanh(\mu_i
H/T)d\mu_i$ where the distribution function $F(\mu_i)$ can be
expressed in terms of the distribution function of cluster sizes.
This mechanism may also contribute to the fractional moment $\mu =
\int_0^\infty \mu_iF(\mu_i)d\mu_i$ obtained from the fit in Fig. 4.
AF clusters of Fe$^{3+}$ around oxygen vacancies were reveal by the
Mossbauer spectroscopy of Sm0$_{1-x}$F$_x$FeAs \cite{mosb2}.

In conclusion, doped Nd(O,F)FeAs can be regarded as a paramagnetic
superconductor in which magnetization is dominated by the
Curie-Weiss paramagnetism of the rare earth ions. Our high field
magnetization measurements have shown no long-range
antiferromagnetic order below $T_c$ despite noticeable AF
correlations above $T_c$.

The work at NHMFL was supported by the NSF grant DMR-0084173 with
support from the state of Florida and AFOSR. CT is grateful to Eun Sang Choi 
for help with high-field measurements.


\end{document}